\documentclass[epj,nopacs]{svjour}
\usepackage{graphicx}
\usepackage{amsmath}
\usepackage{amssymb}
\usepackage{subfigure}

\newcommand{\ups}{{\ensuremath{\Upsilon}}}

\newcommand{\eqs}[1]{\begin{equation} \begin{split} #1\end{split} \end{equation} }

\newcommand{\eg}{{e.g.}}
\newcommand{\etal}{{\it et al.}}

\newcommand{\cf}[1]{{Fig.~\ref{#1}}}

\providecommand{\pp}{{\it pp}}
\providecommand{\pA}{{\it pA}}

\providecommand{\AaAa}{{\it AA}}

\begin{document}
\title{On the mechanisms of heavy-quarkonium hadroproduction}
\author{J.P. Lansberg\inst{1}
\thanks{Present address}\fnmsep\inst{2}%
}                     
\institute{SLAC National Accelerator Laboratory, Stanford University, Stanford, CA 94309, USA
\and
Institut f\"ur Theoretische Physik, Universit\"at Heidelberg, Philosophenweg 19,
D-69120 Heidelberg, Germany \\ \email{lansberg@slac.stanford.edu}}
%
\date{}
%
\abstract{
We discuss the various mechanisms potentially at work in hadroproduction of  
heavy quarkonia in the light of computations of higher-order QCD corrections 
both in the Colour-Singlet (CS) and Colour-Octet (CO) channels and the 
inclusion of the contribution arising from the $s$-channel cut in the CS 
channel. We also discuss new observables meant to better discriminate  between 
these different mechanisms.
\PACS{
      {PACS-key}{discribing text of that key}   \and
      {PACS-key}{discribing text of that key}
     } 
} 
\maketitle
\section{Introduction}
\label{intro}

Heavy quarkonia  are among the most studied bound quark systems. 
It is however fair to say that, for the time being, there is no consensus 
concerning which mechanisms are effectively at work in their production in 
$pp$ collisions, that is at the Tevatron and RHIC. By extension,
available theoretical predictions for the production rates at the  
LHC are rather uncertain. For recent reviews, the reader is guided 
to~\cite{Lansberg:2006dh,Brambilla:2004wf,Kramer:2001hh}
along with some perspectives for the LHC \cite{Lansberg:2008zm}.

To what concerns the $\Upsilon$ states, the latest NLO 
predictions~\cite{Artoisenet:2008fc} in the QCD-based approach of the 
Colour-Singlet Model (CSM)~\cite{CSM_hadron} including for the first time 
some of the important NNLO $\alpha_S^5$ corrections show a
satisfactory agreement with the data coming from the 
Tevatron~\cite{Acosta:2001gv,Abazov:2005yc}. 

On the other hand, to what concerns the charmonium family, it is well-known 
that the first measurements by the CDF Collaboration of the {\it direct}
production of $J/\psi$ and $\psi'$ at $\sqrt{s}=1.8$ 
TeV~\cite{Abe:1997jz,Abe:1997yz} brought to light a striking puzzle. 
They indeed found much larger rates than the prediction of the CSM.

Since then, other approaches were proposed (\eg~the Colour-Octet Mechanism 
(COM) from NRQCD~\cite{Bodwin:1994jh}) or revived (\eg~the Colour-Evaporation 
Model (CEM)~\cite{CEM}). Those are unfortunately not able to reproduce in 
a consistent way experimental studies of both cross-section and polarisation
measurements for charmonia at the 
Tevatron \cite{Abulencia:2007us,Affolder:2000nn} along with the cross 
sections measured by PHENIX at RHIC~\cite{Adare:2006kf}. For instance, 
the seemingly solid prediction of the COM for a transverse polarisation 
of $\psi$'s produced at high transverse momentum is clearly challenged by 
the experimental measurements.  The most natural interpretation of such 
a failure of NRQCD is that the charmonium system is too light for 
relativistic effects to be neglected and  that the quark-velocity expansion 
($v$) of NRQCD~\cite{Bodwin:1994jh} may not be applicable for the  
rather ``light'' $c\bar c$ system. This might be indeed the case in view 
of the aforementioned agreement between theory and the available experimental 
data on production in \pp\ (and inclusive decays) of the significantly heavier 
$\Upsilon$. In this case, relativistic 
corrections are expected to be less important  and the leading state
in the Fock expansion, i.e.  the heavy-quark pair in a colour singlet
$^3S_1$, to be dominant. This would in turn explain why a computation 
incorporating the sole CS channel (LO contribution in the $v$-expansion 
of NRQCD) 
is sufficient -- when $P_T^{-4}$ contributions are considered though.

Recently, many new theoretical results became available. Some completed
our knowledge of quarkonium production like the up-to-date proof~\cite{nayak1} 
of NRQCD factorisation holding true at any order in $v$ in the 
gluon-fragmen\-tation channel -- therefore relevant at large $P_T$ where this
channel should dominate -- and the QCD corrections which we shall
discuss in the next section. Others mainly questioned our understanding
of the mechanisms at work in heavy-quarkonium production.
Firstly, a complete survey of fixed-target
measurements \cite{Maltoni:2006yp} showed that the universality of
the Long Distance Matrix Elements (LDMEs) of NRQCD\footnote{Let us remember
that the universality of the LDMEs is clearly challenged by HERA data on photoproduction of $J/\psi$~\cite{Kramer:2001hh}.} 
cannot certainly be extended to the description
of low-$P_T$ data. Secondly, NRQCD factorisation was shown to require
modifications in fragmentation regions where 3 heavy quarks have similar
momenta \cite{nayak2}. Thirdly, the $c$-quark fragmentation approximation
was shown~\cite{Artoisenet:2007xi} to be only valid at much higher $P_T$ than
expected from the pioneering works~\cite{frag_CSM}.

Parallel to that, we investigated in~\cite{Haberzettl:2007kj,Lansberg:2005pc}
 on the cut contributions due to off-shell and non-static quarks. In 
particular, we questioned the assumption of the CSM\footnote{Note on the 
way that this assumption is also present in NRQCD.} that takes the
heavy quarks forming the quarkonium ($\mathcal{Q}$) as being
on-shell~\cite{CSM_hadron}. If they are not, the usual $s$-channel cut
contributes to the imaginary part of the amplitude and need to be considered on
the same footing as the CSM cut. A first evaluation~\cite{Haberzettl:2007kj} 
of the latter incorporating constraints for the low- and large-$P_T$ (the 
scaling limit) region give rates significantly larger than the usual CSM cut.
 Moreover, low-$P_T$ data from RHIC are very well described without need of 
re-summing initial-gluon contributions. However, as 
expected~\cite{Haberzettl:2007kj}, this approach underestimates the 
cross-section at large values of $P_T$ and other mechanisms have to be 
considered in this region.

In section 2, we present the latest results available on QCD corrections 
to hadroproduction of $J/\psi$, $\psi'$ and $\Upsilon(nS)$. In section 3, 
we discuss how the $s$-channel cut contribution to the CS channel can be 
evaluated and we present a comparison with data. In section 4, we briefly 
review other recent theoretical results. In section 5, we show how the study of 
the production of quarkonia in association with a heavy-quark pair of the same 
flavour may be used to disentangle between the different mechanisms proposed
to explain quarkonium production. Finally, we present our conclusions and
 outlooks.

\section{QCD corrections}

More than ten years ago now, the very first NLO calculation on quarkonium 
production to date became available. It was  centred on unpolarised 
photoproduction of $\psi$~\cite{Kramer:1995nb} via a colour-singlet (CS)
transition. Later on, NLO corrections were computed for direct $\gamma\gamma$
collisions~\cite{Klasen:2004az,Klasen:2004tz} for which it had been previously
shown~\cite{Klasen:2001cu} that the LO CS contribution alone was not able 
to correctly reproduce the measured rates by DELPHI~\cite{Abdallah:2003du}. 
NLO corrections have also recently been computed for the integrated cross 
section of two $J/\psi$-production observables at the $B$-factories: 
$J/\psi + c \bar c$~\cite{Zhang:2006ay} and 
$J/\psi+\eta_c$~\cite{Zhang:2005cha}. As of today, only the full 
colour-octet (CO) contributions to direct $\gamma\gamma$ collisions have 
been evaluated at NLO for $P_T>0$~\cite{Klasen:2004az,Klasen:2004tz}. 

At the LHC and the Tevatron, $\psi$ and $\Upsilon$ production proceeds
most uniquely via gluon-fusion processes. The corresponding cross
section at NLO ($\alpha_S^4$ for hadroproduction processes) are
significantly more complicated to compute than the former ones 
and became only available one year 
ago~\cite{Campbell:2007ws,Artoisenet:2007xi}. We shall discuss them in the 
next section.

The common feature of all these calculations is the significant size of
the NLO corrections, in particular for large transverse momenta  $P_T$ of the
quarkonia for the computations of differential cross sections in $P_T$. In 
$\gamma p$ an \pp~collisions, QCD corrections to the CS production indeed 
open  new channels with a different behaviour in $P_T$ which  raise 
substantially the cross section in the large-$P_T$ region. 

Let us discuss this shortly for the gluon-fusion processes which dominate the 
yield in $pp$.  If we only take into account the CS transition to $^3S_1$ 
quarkonia, it is well known that the differential cross section at LO as a 
function of $P_T$ scale like $P^{-8}_T$ \cite{CSM_hadron}. This is expected from
contributions coming from the typical ``box'' graphs of \cf{diagrams} (a). 
At NLO~\cite{Campbell:2007ws,Artoisenet:2007xi}, we can distinguish three 
noticeable classes of contributions. First, we have the loop contributions 
as shown on \cf{diagrams} (b), which are UV divergent\footnote{These 
divergences  can be treated as usual using dimensional regularisation, 
see \eg~\cite{Campbell:2007ws}.} but as far their $P_T$ scaling is concerned, 
they would still scale like  $P^{-8}_T$. Then we have the $t$-channel gluon 
exchange graphs like on  \cf{diagrams} (c). They scale like $P^{-6}_T$. For 
sufficiently large $P_T$, their smoother $P_T$ behaviour can easily compensate
 their $\alpha_S$ suppression compared to the LO ($\alpha_S^3$) contributions.
They are therefore expected to dominate over the whole set of diagrams up to 
$\alpha_S^4$. To be complete, we should not forget the $\alpha_S^4$ 
contributions from ${\cal Q} + Q \bar Q$ (where $Q$ is of the same flavour 
as the quarks in $\cal Q$). Indeed, one subset of graphs for 
${\cal Q} + Q \bar Q$ is fragmentation-like  (see \cf{diagrams} (d)) and 
scales like $P^{-4}_T$. Such contributions are therefore expected to dominate 
at large $P_T$, where the smoother decrease in $P_T$ is enough to compensate 
the suppression in $\alpha_S$ and the one due to the production of 4 heavy 
quarks. As mentioned above, in practice~\cite{Artoisenet:2007xi}, this happens
 at larger $P_T$ than as expected before~\cite{frag_CSM}. We shall come back 
to this channel later. In the next sections, we shall discuss the impact of 
the NLO corrections to the CS channels and then a first computation including 
the {\it a priori} dominant $\alpha_S^5$ contributions i.e. topologies
 illustrated by \cf{diagrams} (e) and (f).

\begin{figure}[ht!]
\centering
\subfigure[]{\includegraphics[scale=.33]{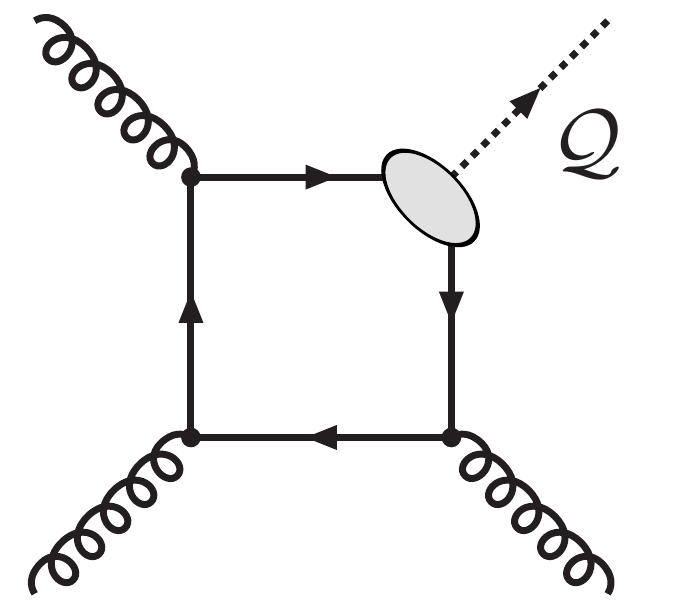}}
\subfigure[]{\includegraphics[scale=.33]{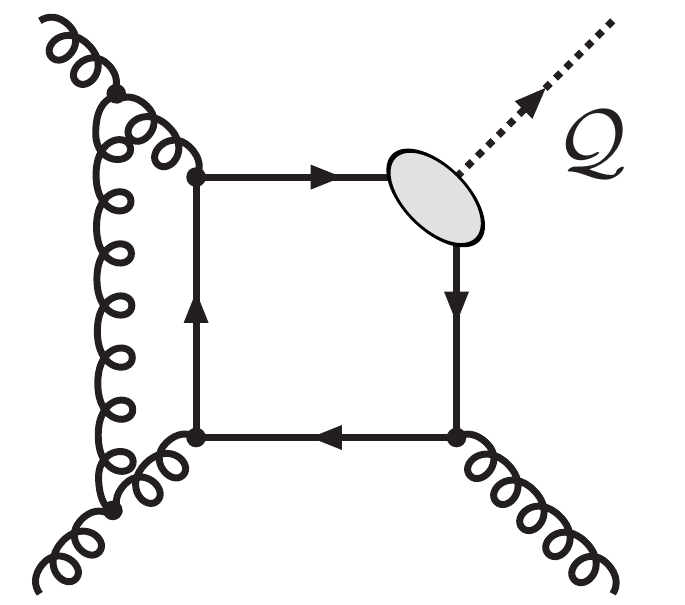}}
\subfigure[]{\includegraphics[scale=.33]{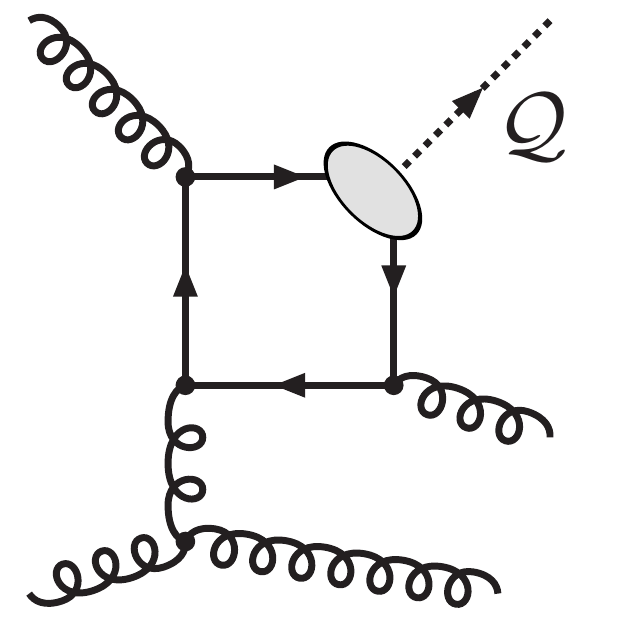}}
\subfigure[]{\includegraphics[scale=.33]{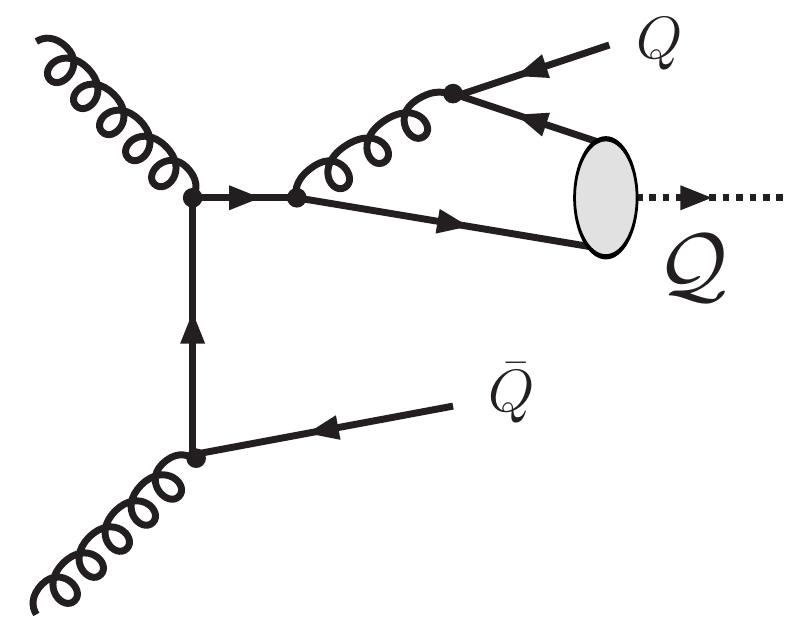}}
\subfigure[]{\includegraphics[scale=.33]{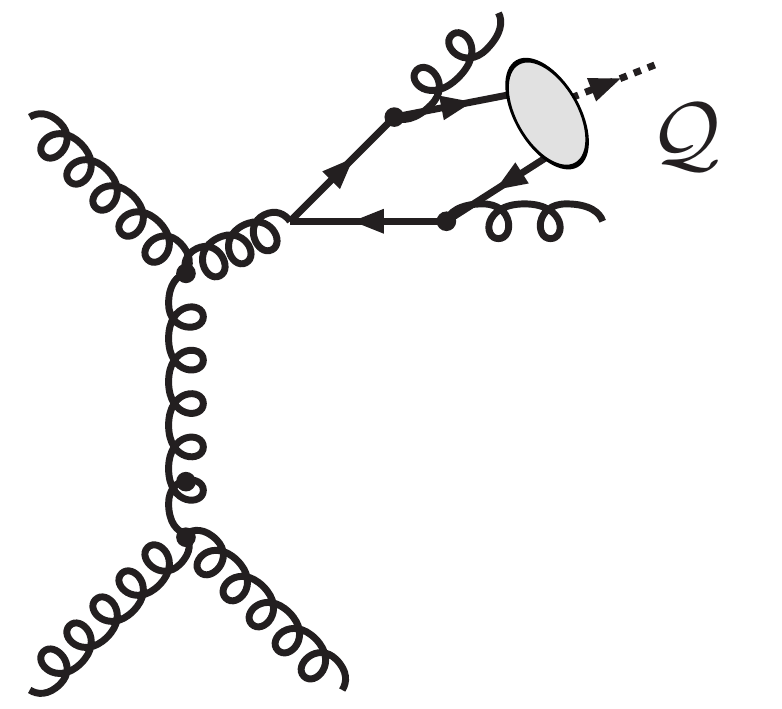}}
\subfigure[]{\includegraphics[scale=.33]{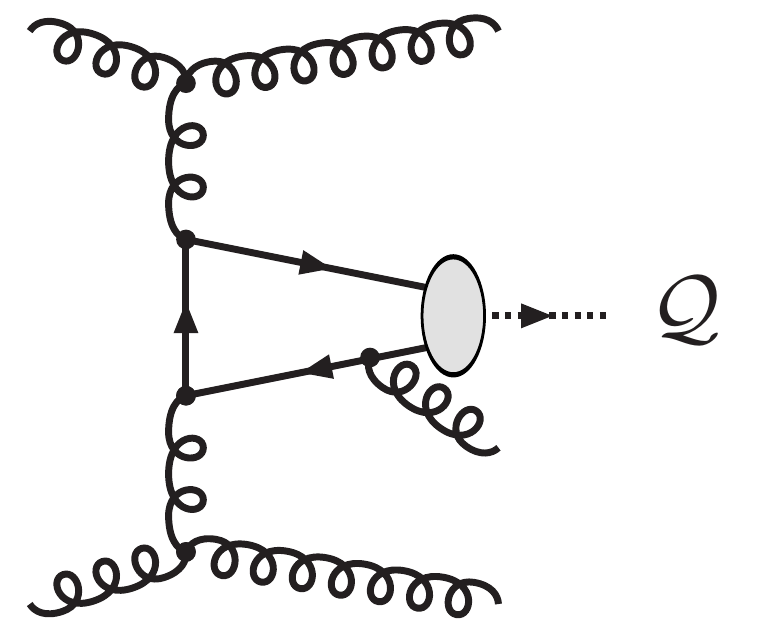}}
\subfigure[]{\includegraphics[scale=.33]{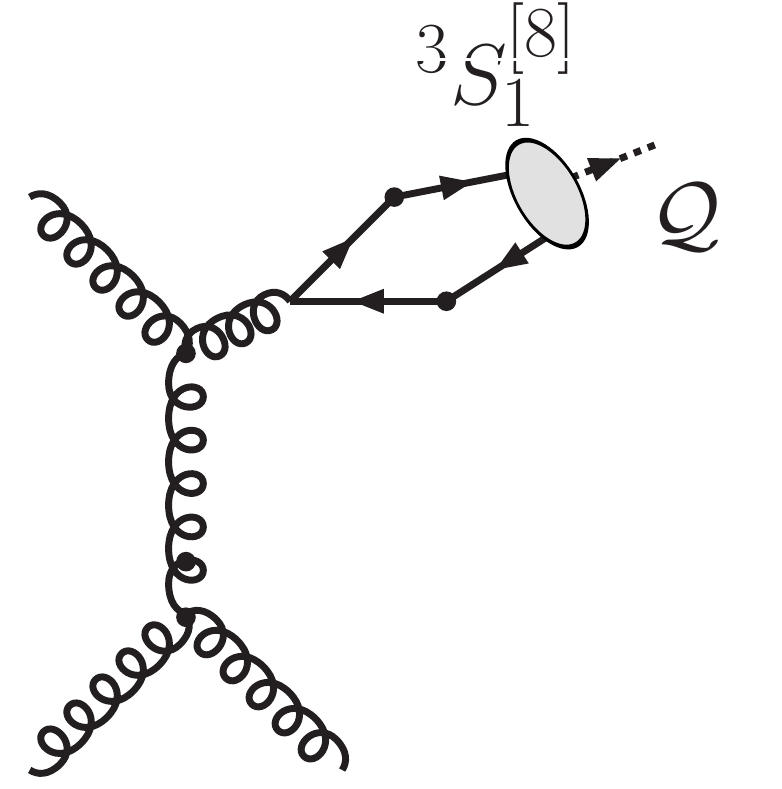}}
\subfigure[]{\includegraphics[scale=.33]{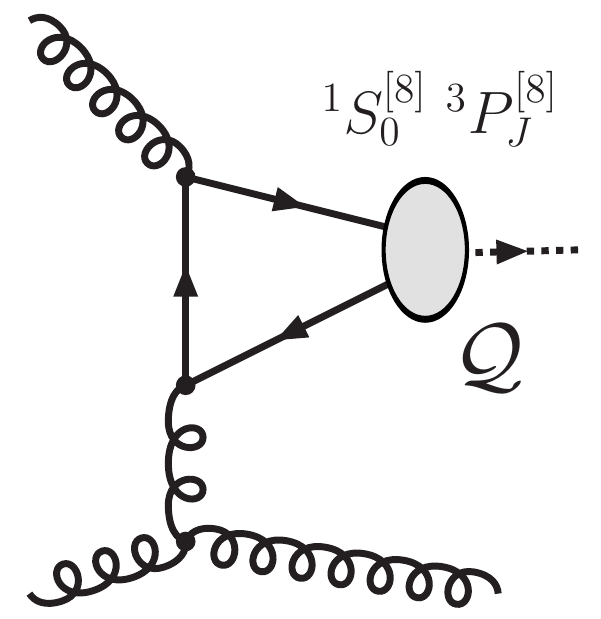}}
\caption{Representative diagrams contributing to $^3S_1$ hadroproduction via 
Colour-Singlet channels at orders $\alpha_S^3$ (a), $\alpha_S^4$ (b,c,d), 
$\alpha_S^5$ (e,f) and via Colour-Octet channels at orders $\alpha_S^3$ (g,h). 
The quark and antiquark attached to the ellipsis are taken as on-shell
and their relative velocity $v$ is set to zero.}
\label{diagrams}
\end{figure}

To what concerns the CO contributions, the effects of NLO (here $\alpha_S^4$)
contributions are expected to be milder. Indeed, the contributions from CO 
transitions are {\it a priori} suppressed by $v^4$ and the reason why they 
can still appear significant comes from a lower power in $\alpha_S$ for 
similar topologies (and thus similar $P_T$ scaling). For instance, compare 
\cf{diagrams} (g) to  \cf{diagrams} (e) and  \cf{diagrams} (h) to  
\cf{diagrams} (c). At LO, $P_T^{-6}$ and $P_T^{-4}$ scaling are therefore 
already present. As a result, including $\alpha_S^4$ contributions will not 
open any new channels and NLO corrections are expected to be described by a 
roughly constant $K$ factor. As we shall see,  this is indeed the trend seen 
in the results of~\cite{Gong:2008ft} in which the NLO corrections to 
$^1S_0^{[8]}$  and $^3S_1^{[8]}$ colour octets going to $J/\psi$ were considered.

\subsection{NLO corrections for Colour-Singlet channels }

Let us first present a comparison between the measurements by the CDF 
collaboration and the result for the $J/\psi$ obtained following the 
procedure explained in~\cite{Campbell:2007ws,Artoisenet:2007xi}. 
It is worth noting that the $\chi_c$ cross sections are not available for 
now at NLO accuracy. This would be necessary if we wanted to predict at this 
accuracy prompt-$J/\psi$ production cross section, in order to compare with 
the most recent measurements of RUN II~\cite{Acosta:2004yw}. These focused
only on the prompt yield. As a makeshift, we have multiplied those data by
 the averaged fraction of direct  $J/\psi$ measured during RUN 
I~\cite{Abe:1997yz} for the rather similar beam energy 1.8 TeV and a similar 
range in $P_T$ and rapidity: $\langle F^{direct}\rangle=64\pm 6 \%$.

In our calculation, we set $m_c=1.5\pm 0.1$ GeV and $m_b=4.75\pm 0.25$ GeV. 
We used the PDF set CTEQ6L1 (resp. CTEQ6\_M) \cite{Pumplin:2002vw} for LO 
(resp. NLO) cross sections, and always kept the factorisation scale equal 
to the renormalisation scale: $\mu_f=\mu_r$. Except for the associated 
production channel where we took $\mu_0=\sqrt{(2 m_{\cal Q})^2+ P_T^2}$,
 the central scale is fixed at $\mu_0=\sqrt{m_{\cal Q}^2+ P_T^2}$ and then 
was varied by a factor 
of 2. To what concerns the non-perturbative inputs, we used the values related
 to the BT potential~\cite{Eichten:1995ch} : 
$\langle \mathcal{O}^{J/\psi} (^3 S^{[1]}_1  ) \rangle =1.16$ GeV$^3$ 
and $\langle \mathcal{O}^{\Upsilon}(^3 S^{[1]}_1  ) \rangle=9.28$ GeV$^3$.
For $\Upsilon(1S)$ production, we considered the prompt measurement at 
$\sqrt{s}=1.8$ TeV in~\cite{Acosta:2001gv}, multiplied by the averaged 
direct fraction obtained in~\cite{Affolder:1999wm}: 
$\langle F^{direct}\rangle=50\pm 12 \%$.

\begin{figure}[h!]
\subfigure[$J/\psi$]{\centerline{\includegraphics[width=.8\columnwidth]{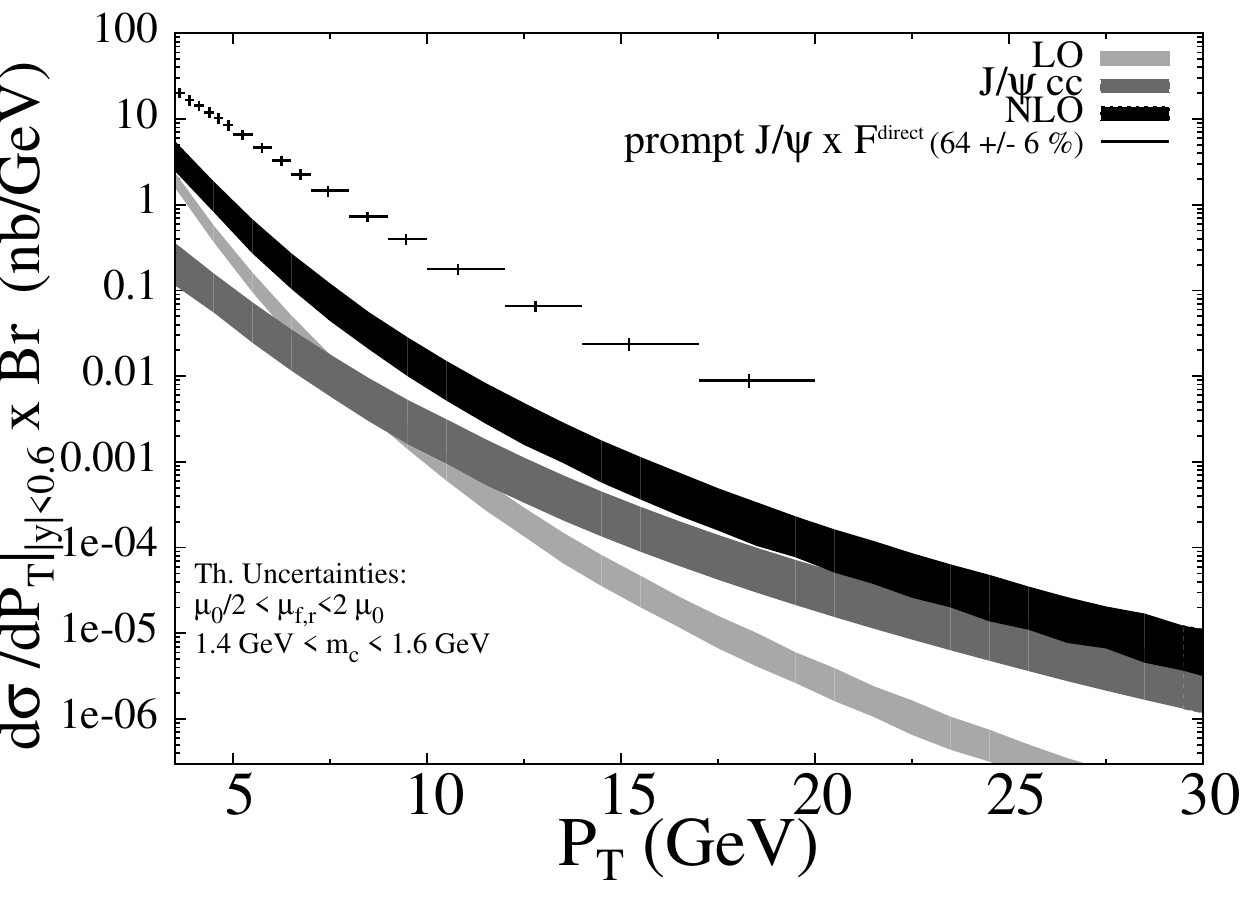}}}
\subfigure[ $\Upsilon(1S)$]{\centerline{\includegraphics[width=\columnwidth]{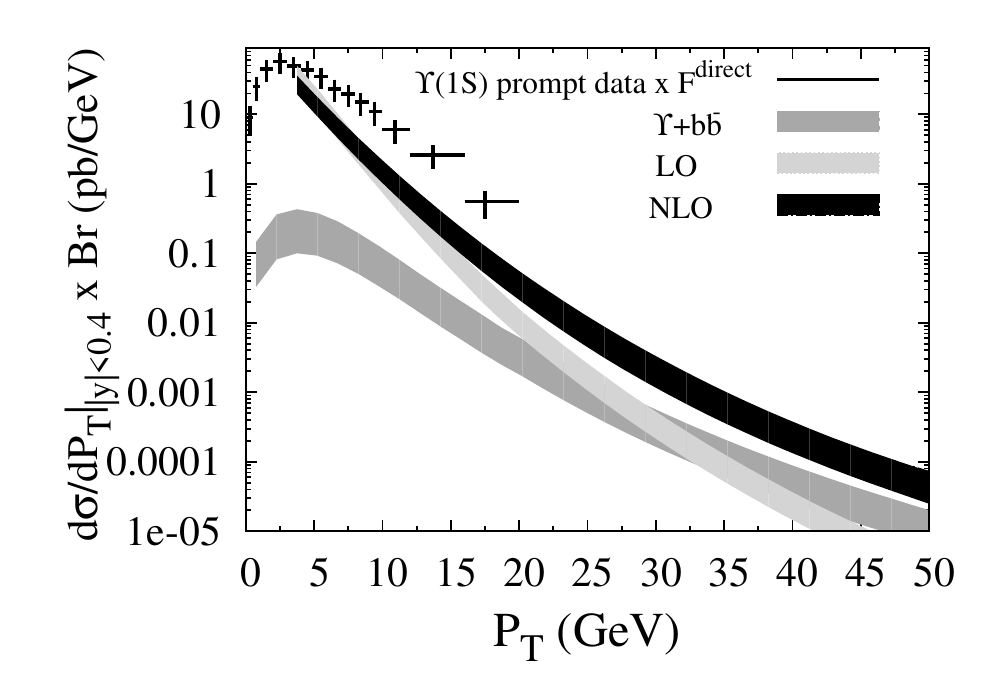}}}
\caption{Differential cross sections at 
NLO accuracy as function of the quarkonium transverse momentum $P_T$ at the Tevatron ($\sqrt{s}=1.96$ TeV).} 
\label{fig:dsigNLO}
\end{figure} 

In both case, we have an illustration of the previous discussion. The 
differential cross section for the LO contribution, i.e. $gg \to J/\psi g$, 
has the steepest slope and is already an order of magnitude smaller than the 
NLO contribution at $P_T\simeq 10$ GeV.  The differential cross section for 
${\cal Q}+ Q\bar Q$ has the smoothest slope. In the case of $J/\psi$, it 
starts to be significant for $P_T> 20$ GeV. For the $\Upsilon$, the 
suppression due to the production of 4 $b$ quark is stronger  and this yield 
remains negligible in the accessible value of $P_T$. The bands denoted NLO 
refer to all the contributions up to order $\alpha_S^4$.

  Those results
were recently confirmed in~\cite{Gong:2008sn,Gong:2008hk}. In the latter 
papers, the polarisation information was kept and the observable $\alpha$ was 
also computed. However, it is important to stress that for $\psi$ and 
$\Upsilon$ production the CS yields predicted at the NLO accuracy are still 
clearly below the experimental data especially at large $P_T$. In this respect,
the predictions for the polarisation at this order cannot be usefully compared
 to the data.

 The conclusion is that, in general, the inclusion of NLO contributions
bring the CS predictions considerably closer to the data, although agreement 
is only reached at NLO in the photoproduction case~\cite{Kramer:1995nb}.

\begin{figure*}[ht!]\centering
\subfigure{\includegraphics[width=.9\columnwidth]{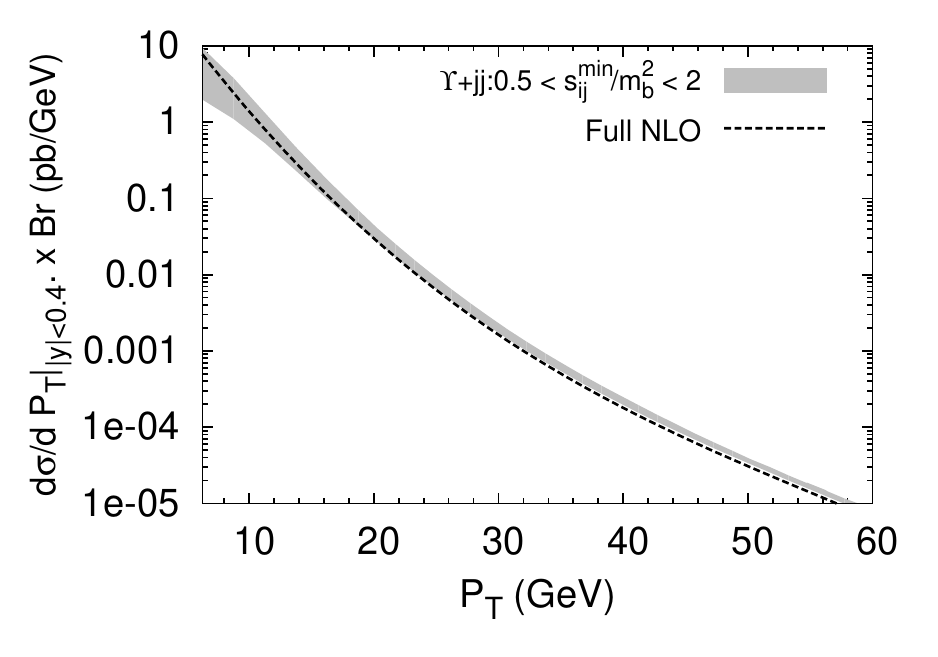}}
\subfigure{\includegraphics[width=.9\columnwidth]{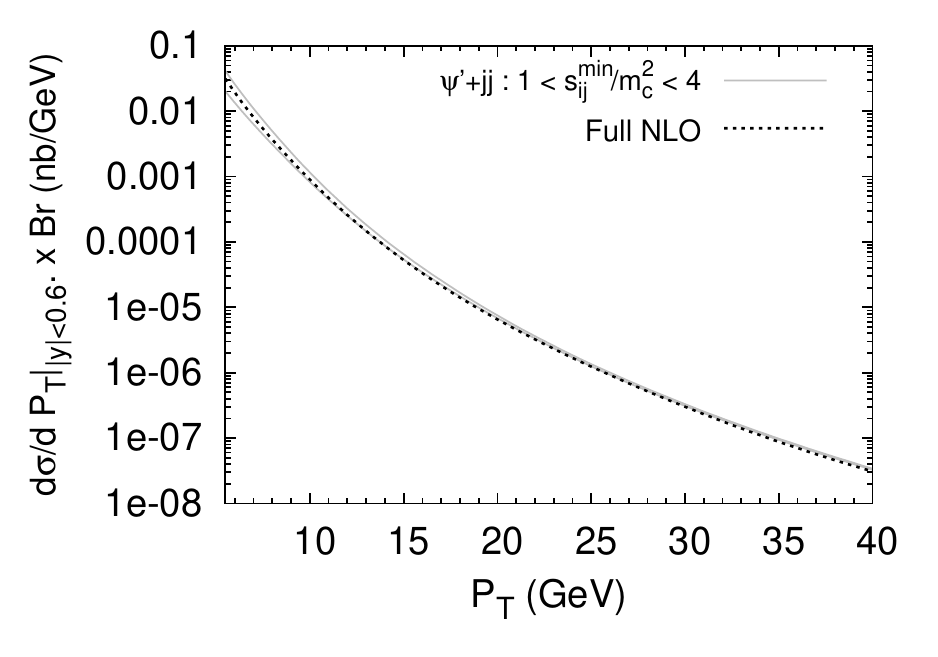}}
\caption{Full computation at NLO for (Left) $\Upsilon(1S) + X$ (dashed line)  
vs. $\Upsilon(1S)$ + 2 light partons with a cut on $s_{ij}^{\rm min}$ (grey band),
(Right) $\psi(2S) + X$ (dashed line)  vs. $\psi(2S)$ + 2 light partons with a
 cut on $s_{ij}^{\rm min}$ (grey curves)} \label{fig:NLO-2jet}
\end{figure*}

\subsection{NLO corrections for Colour-Octet channels}

As aforementioned, NRQCD has reached a certain success by explaining 
the main features of charmonium and bottomonium 
hadroproduction via the introduction of the Colour-Octet (CO) mechanism. 
It indeed provides a good description of the $P_T$-differential cross-section 
for the direct~$J/\psi$ and $\psi'$ for $P_T\gtrsim 5$~GeV as measured by CDF 
in $p\bar p$~\cite{Abe:1997jz,Abe:1997yz}. A reasonable agreement was also
obtained with  the first PHENIX measurements in \pp~at $\sqrt{s}=200$ 
GeV~\cite{Adler:2003qs,Cooper:2004qe}. In both cases, the cross-section is 
dominated by the gluon fragmentation into a colour-octet $S$-wave state. 
Following the heavy-quark spin symmetry~\cite{Bodwin:1994jh} of NRQCD, the 
latter mechanism leads to transversally polarised $J/\psi$ and $\psi'$, the 
parent fragmentating gluon being mostly on-shell and thus transversally 
polarised at high $P_T$.

However,   $J/\psi$ and $\psi'$ are not seen to be transverse by the CDF
experiment~\cite{Abulencia:2007us}. It measured a slight
longitudinal polarisation for both the prompt~$J/\psi$ and
direct~$\psi'$ yield.  It is worth noting here that the feed-down from
$\chi_c$ can influence significantly the polarisation of the prompt
$J/\psi$ yield -- this was taken into account in the NRQCD-based
predictions~\cite{Braaten:1999qk}.
Moreover, the recent preliminary result
from PHENIX~\cite{ErmiasHP08} indicates a polarisation compatible with
zero for the total $J/\psi$ production at forward rapidity ($1.2 < |y|
< 2.2$), but with large uncertainties.

Very recently, 
CO contributions from $S$ waves ($^1S_0^{[8]}$ and $^3S_1^{[8]}$) have
become available~\cite{Gong:2008ft} for hadroproduction. A complete
phenomenological study is not yet available though. Anyhow this 
confirms that NLO corrections do not affect significantly the $P_T$ 
dependence as expected from the introductory discussion of this section.

Let us define  $K$ factors as the ratios of NLO to LO  cross section 
for a given CO channel. For the Tevatron, they are about 1.2 for the 
$^1S_0^{[8]}$ state and 1.1 for the $^3S_1^{[8]}$ (at the LHC, they are both 
about 0.8). Consequently, the value of the CO Long Distance Matrix Elements 
(LDMEs) fit to the Tevatron data at LO
$\langle O^{J/\psi} \big( ^3S_1^{[8]}\big) \rangle \simeq0.0012$ GeV$^3$ and
$\langle O^{J/\psi} \big( ^1S_0^{[8]}\big) \rangle\simeq 0.0045$
GeV$^3$~\cite{Kramer:2001hh} would be at most reduced by
15\%. In this respect,  the NLO corrections to the octets do not improve
the universality of the matrix elements when the idea of the dominance of
the CO transitions is confronted to the data on photoproduction from HERA.

According to the author of~\cite{Gong:2008ft}, it is not possible to obtain a 
satisfactory $P_T$ distribution in terms of a unique $\langle O^H_n \rangle$ 
value when considering the whole range in $P_T$ analysed by CDF. More 
precisely, they did not consider the experimental data with $P_T < 6$~GeV, for
 which it seems that other mechanisms have to be at work if we believe that 
the COM is responsible for the major part of the cross section at large $P_T$.

This in any case emphasises the need for more  work dedicated  to the 
description of the low-$P_T$ region and maybe the relevance of the study of 
$s$-channel cut contribution, which we discuss later.  Last but not least, 
the polarisation from CO transitions appears not to be modified at NLO with 
respect to LO results. Overall, this recent first study of CO contributions at 
NLO in hadroproduction at $P_T>0$ sounds like a confirmation of the flagrant 
discrepancy between the NRQCD predictions for the polarisation of the $J/\psi$
 and the experimental measurements from the CDF 
collaboration~\cite{Abulencia:2007us}.

\subsection{QCD corrections  up to $\alpha^5_S$}

As noted above, the discrepancy between the NLO computations 
for the CSM and the 
experimental data, both for $\psi$ and $\Upsilon$ still grows with $P_T$. If 
we parallel that to the existence of new $P_T^{-4}$ channel at order 
$\alpha^5_S$, it is reasonable to wonder what their size are effectively . 

In fact, their contributions can be evaluated in a relatively 
``simple''\footnote{``simple'' compared to a full  --out-of-reach-- NNLO 
computation and thanks to the automated generator of matrix elements 
MadOnia~\cite{Artoisenet:2007qm}.} and reliable way by computing the 
$\alpha^5_S$ contributions consisting in the production of a $\cal Q$ with 3 
light partons (noted $j$ thereafter). Among them are the topologies 
of~\cf{diagrams} (d) (gluon fragmentation) and~\cf{diagrams} (e) 
(``high-energy enhanced''), these close the list of kinematical enhancements 
from higher-order QCD corrections. This $\alpha^5_S$ subset being the LO for 
a physical process ($pp \to {\cal Q} +jjj$), its contribution is finite except
 for soft and collinear divergences.

To avoid such divergences, we impose a lower bound on the invariant-mass of 
any light partons ($s_{ij}$). For the new channels opening up at $\alpha^5_S$, 
and which specifically interest us, the dependence on this cut is to get 
smaller for large $P_T$ since no collinear or soft divergences can appear there.
For other channels, whose LO contribution is at $\alpha^3_S$ or $\alpha^4_S$,
the cut would produce logarithms of $s_{ij}/s_{ij}^{\rm min}$. Those can be 
large. Nevertheless, they can be factorised over their corresponding LO 
contribution, which scales at most as $P_T^{-6}$. The sensitivity on 
$s_{ij}^{\rm min}$ is thus expected to come to nothing at large $P_T$.

Thanks to the exact NLO computation of~\cite{Campbell:2007ws}, such a 
procedure can be tested for the process $pp \to {\cal Q} +jj$. For instance, 
the differential cross section for the real $\alpha_s^4$ corrections, 
$\Upsilon(1S)+jj$ production, is displayed in \cf{fig:NLO-2jet} (Left). The 
grey band illustrates the sensitivity to the invariant-mass cut 
$s_{ij}^{\rm min}$ between any pairs of light partons when it is varied from 
$0.5 m_b^2$ to $2m_b^2$.  The yield becomes insensitive to the value of 
$s_{ij}^{\rm min}$ as $P_T$ increases, and it reproduces very accurately the 
differential cross section at NLO accuracy. In the charmonium case, the 
similar contributions from $pp \to \psi' +jj$ matches even better, for lower
 $P_T$ and with a smaller dependence of $s_{ij}^{\rm min}$ the full NLO 
computation, as seen on \cf{fig:NLO-2jet} (Right).

\begin{figure}[ht!]\centering
\subfigure[]{\includegraphics[width=.9\columnwidth]{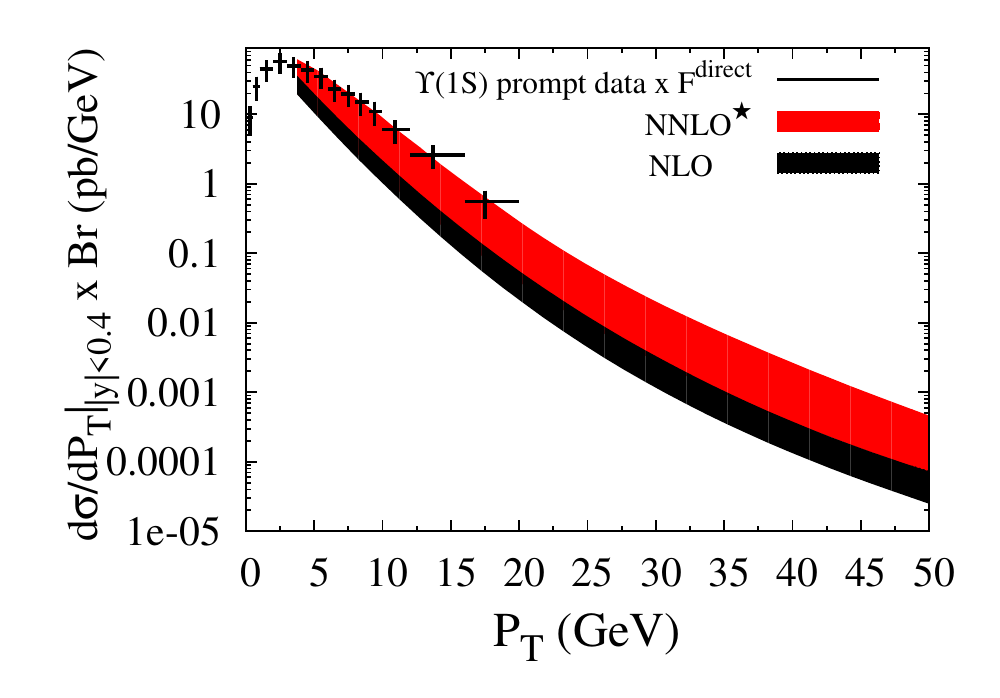}}
\subfigure[]{\includegraphics[width=.9\columnwidth]{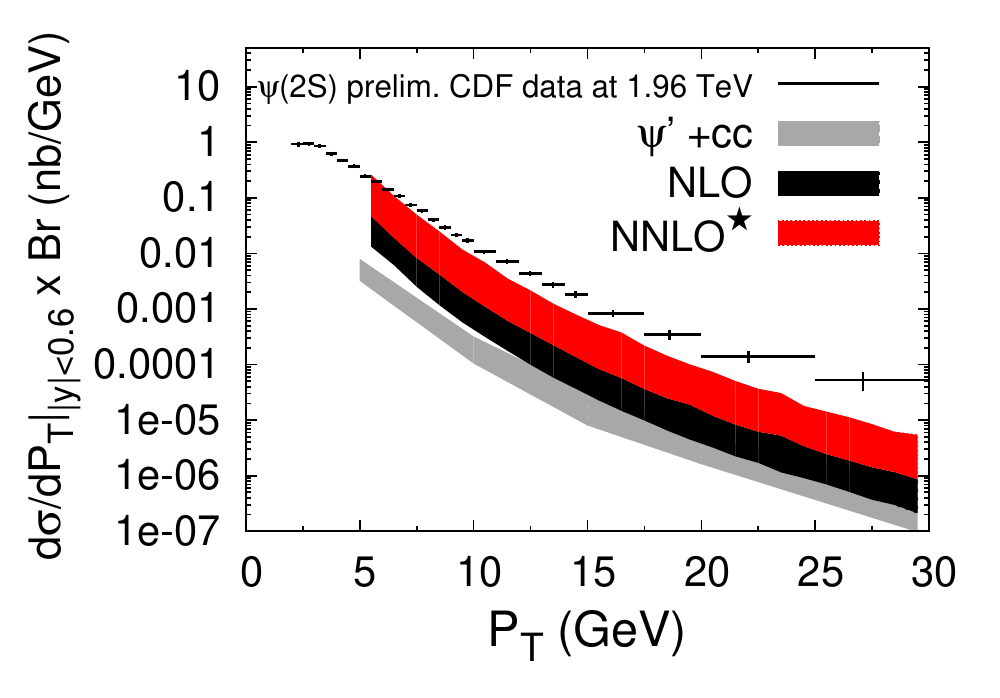}}
\caption{Comparison between differential cross sections at 
NLO and NNLO$^\star$ accuracy as function  as function of the $\cal Q$ 
transverse  momentum $P_T$ at the Tevatron ($\sqrt{s}=1.96$ TeV) and the data 
for  (a) $\Upsilon(1S)$~\cite{Acosta:2001gv} \& (b) direct 
$\psi(2S)$~\cite{cdf-psi2s-prelim}.} \label{fig:dsdpt_NNLO}
\end{figure}

We now turn to the results concerning the real contribution  at $\alpha^5_S$,
which we refer to as NNLO$^\star$. We used the approach described in 
Ref.~\cite{Artoisenet:2007qm}, which allows the automatic generation of both 
the subprocesses and  the corresponding scattering amplitudes.
The differential cross-sections for $\Upsilon(1S)$ and $\psi(2S)$
are shown in~\cf{fig:dsdpt_NNLO}.  The red band (referred to as
NNLO$^\star$) corresponds to the sum of the NLO yield and the
${\cal Q}+jjj$ contributions.  In the $\Upsilon$ case, the contribution 
from $\Upsilon$ with three light partons fills the gap between the data and
the NLO calculation, while for the $\psi(2S)$ there seems to remain a small 
gap between the NNLO$^\star$ 
band and the preliminary CDF data~\cite{cdf-psi2s-prelim}. In both cases, 
the $\alpha_S^5$ contribution is very sensitive
to the choice of the renormalisation scale, $\mu_r$.  This is
expected: for moderate values of the $P_T$, the missing virtual part
might be important, whereas at large $P_T$, the yield is dominated by
Born-level $\alpha_S^5$-channels from which we expect a large
dependence on $\mu_r$. Even though the uncertainty on the
normalisation is rather large, the prediction of the $P_T$ shape is
quite stable and agrees well with the behaviour found in the 
data~\cite{Acosta:2001gv,Abazov:2005yc,cdf-psi2s-prelim}.

\begin{figure}[htb!]\centering
\subfigure[]{\includegraphics[width=.98\columnwidth]{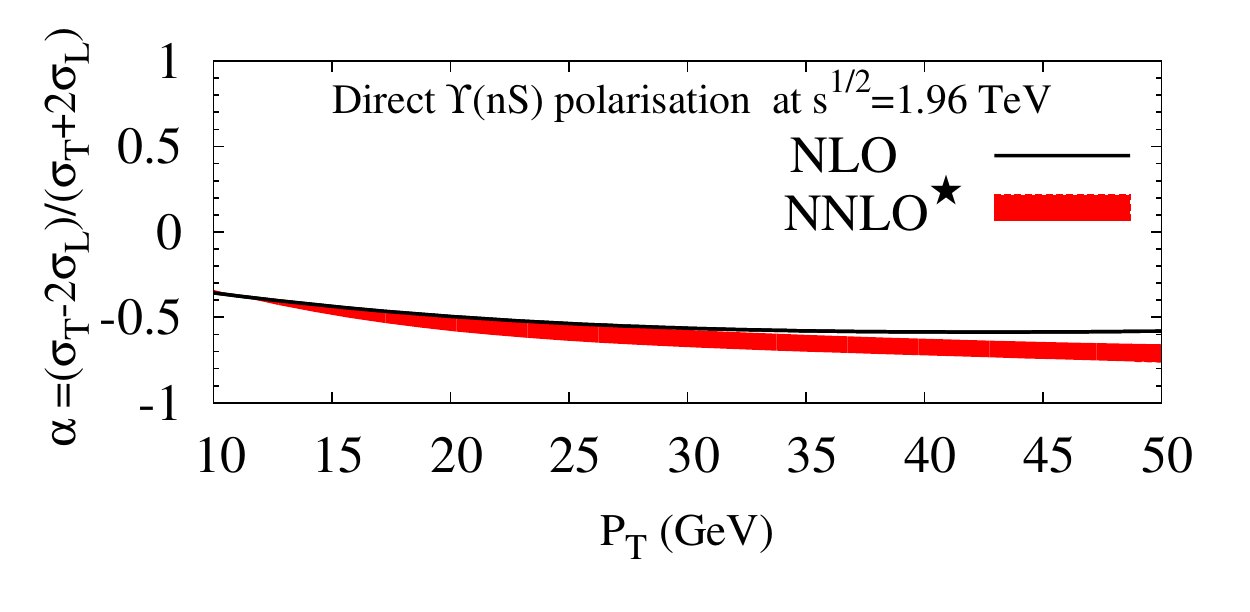}}
\subfigure[]{\includegraphics[width=.98\columnwidth]{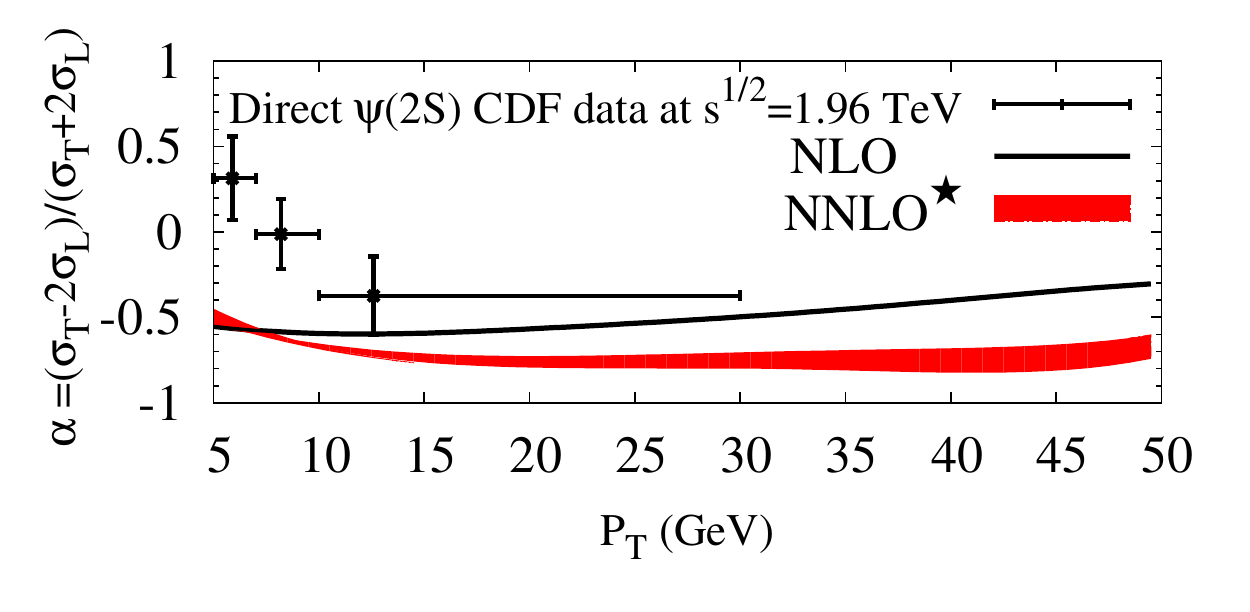}}
\caption{Polarisation of (a) $\Upsilon(nS)$ ((b) $\psi(2S)$)  directly 
produced as function of its transverse momentum $P_T$ at the Tevatron.} 
\label{fig:alpha_NNLO}
\end{figure} 

Concerning the polarisation, the {\it direct} yield is predicted to be mostly
longitudinal, see \cf{fig:alpha_NNLO} (a). However, existing experimental data
 for $\Upsilon$ are centred on the {\it prompt} 
yield~\cite{Acosta:2001gv,D0:2008za}. In order to draw further conclusions, we
 would need first to gain some insights on NLO corrections to $P$-wave 
production at $P_T>0$ . Yet, since the yield from $P$-wave feed-down is likely 
to give transversely polarised $\ups$, the trend is more than encouraging. To 
what concerns $\psi(2S)$, one should be very careful before any comparison 
with experimental measurements since the yield is not exactly reproduced. 
Having this in mind, one sees in~\cf{fig:alpha_NNLO} (b) that the trend for 
longitudinally polarised $\psi(2S)$ is reproduced but more marked. At very 
large $P_T$ where the contribution from $\psi(2S)+c \bar c$ becomes more and 
more significant the polarisation gets slightly less negative. In any case, 
further investigations are needed to draw any conclusions.

\section{$s$-channel cut contribution}

In this section, we briefly review our first evaluation of the $s$-channel 
cut contribution in hadroproduction of 
$J/\psi$~\cite{Haberzettl:2007kj,Lansberg:2005pc} and present outlooks for 
necessary future investigations in this direction.

If the quarks which constitute the $\cal Q$ are not on their mass shell, it is
 not possible to factorise in a gauge invariant way the amplitude for the 
production of those quarks and the one responsible for their binding into the
 $\cal Q$. In other words, if the latter amplitude is given by a (3-point) 
Bethe-Salpeter vertex function, the set of Feynman  diagrams shown 
on~\cf{fig:Jpsibox} (a,b) constructed from this vertex  will not be 
gauge-invariant. Indeed, we have to introduce a 4-point function (or contact 
term) and build up from it  new contributions to the production amplitude 
(\cf{fig:Jpsibox} (c)). Note that  the 4-point function $c\bar{c}J/\psi g$ 
could be interpreted as a dynamical  generalisation of a CO matrix elements for
the transition between a $C=+1$ CO state and a  $J/\psi$.

\begin{figure}[htb!]\centering
  \includegraphics[width=.985\columnwidth,clip=]{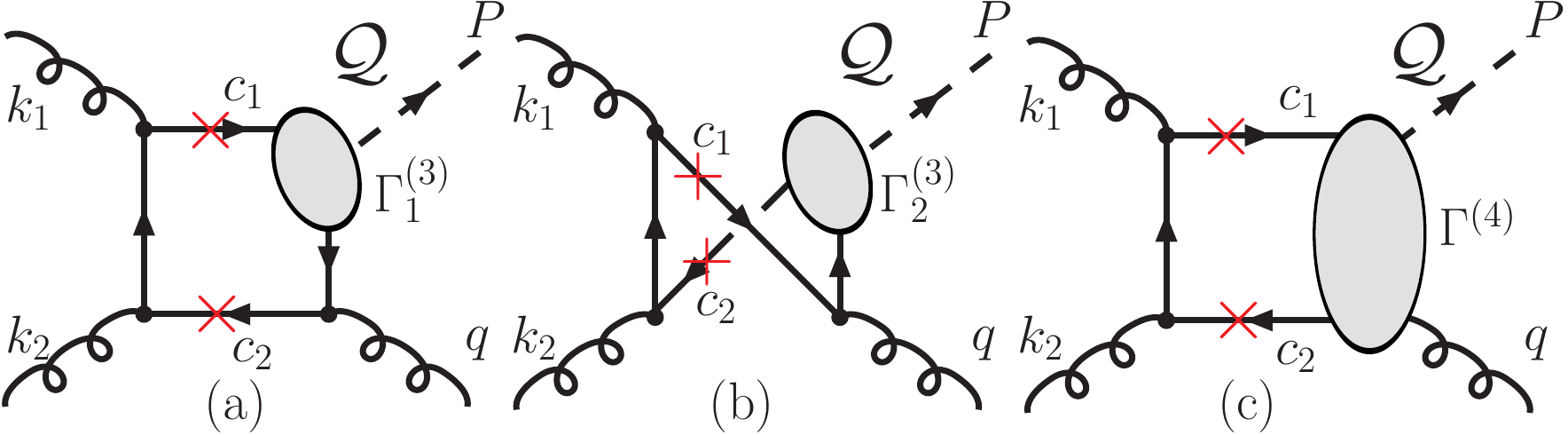}
  \caption{\label{fig:Jpsibox}%
  (a) \& (b) Leading-order (LO) $s$-channel cut diagrams contributing to $gg\to
\mathcal{Q}g$ with direct and crossed box diagrams employing  the
3-point $c\bar{c}\,\cal Q$ vertex.  The crosses indicate that the quarks are 
on-shell. (c) Box diagram with the (4-point) $c\bar{c}\mathcal{Q}g$ contact 
term mandated by gauge invariance. }
\end{figure}

Gauge invariance relates the 4-point function to the 3-point one, but not 
univocally since it does not constrain its transverse contribution with 
respect to the emitted gluon. However there is an elegant way to parametrise 
this freedom by an auxiliary function $F$~\cite{Haberzettl:1997jg,Haberzettl:1998eq,Davidson:2001rk,Haberzettl:2006bn}. To see this,  let us define the 
3-point function 
\begin{equation}\label{vf}
\Gamma^{(3)}_{\mu}(p,P) = \Gamma(p,P) \gamma_\mu~,
\end{equation}
where  $P\equiv p_{1}-p_{2}$  and $p\equiv(p_{1}+p_{2})/2$ are the total and
relative momenta, respectively, of the two quarks bound as a quarkonium state,
with $p_1$ and $p_2$ being their individual four-momenta and the 4-point one by
\begin{equation}
\Gamma^{(4)} =-ig_s T^{a}_{ik} M_c^\nu \gamma^\mu~,
\label{eq:4pointfct}
\end{equation}
 where $g_s$ is the strong coupling constant, $T^{a}_{ik}$ the
colour matrix, and $\mu$ and $\nu$ are the Lorentz indices of the outgoing
$J/\psi$ and gluon, respectively. For simplicity, we have suppressed all
indices on the left-hand side. The $c\bar{c}\,J/\psi$ vertex function
$\Gamma^{(3)}$ with the kinematics of the direct graph is denoted here by
$\Gamma_1$ and for the crossed graph by $\Gamma_2$, {\it i.e.}, $\Gamma_1 =
\Gamma\left(c_1-\frac{P}{2},P\right)$ and $\Gamma_2 =
\Gamma\left(c_2+\frac{P}{2},P\right)$, as shown in Figs.~\ref{fig:Jpsibox}(a)
and (b).

 One easily verifies that any contact current defined as 
follows\footnote{We have taken  $c_1^2=c_2^2=m^2$ and $P^2=M^2$  with $m$ and
$M$ being the masses of the quark and the  $J/\psi$.}
\begin{equation}
M_c^\nu = \frac{(2c_2+q)^\nu\left(\Gamma_1-F\right)}{(c_2+q)^2-m^2}
 +\frac{(2c_1-q)^\nu\left(\Gamma_2-F\right)}{(c_1-q)^2-m^2}~,
 \label{Mc}
\end{equation}
satisfies the gauge-invariance requirement~\cite{Haberzettl:2007kj}
for any value of the function $F=F(c_1,c_2,q)$ . 

Nonetheless, the function $F(c_1,c_2,q)$ must be chosen so that the current
(\ref{Mc}) satisfies crossing symmetry (\textit{i.e.}, symmetry under the 
exchange $c_1\leftrightarrow -c_2$) and is free of singularities. The latter
 constraint implies $F=\Gamma_0$ at either pole position, {\it i.e.}, when 
$(c_2+q)^2=m^2$
or $(c_1-q)^2=m^2$, where the constant $\Gamma_0$ is the (unphysical) value of
the momentum distribution $\Gamma(p,P)$ when all three legs of the vertex are
on their respective mass shells. In principle, employing gauge invariance as
the only constraint, we may take $F=\Gamma_0$ everywhere. This corresponds to
the minimal substitution discussed by Drell and Lee~\cite{Drell:1971vx} (for a
complete derivation see~\cite{Ohta:1989ji}) who pointed out, however, that this
does not provide the correct scaling properties at large energies, which means
within the present context that $F=\Gamma_0$ would not lead to the expected
$P_T$ scaling of the amplitude. See~\cite{Lansberg:2008jn} for a numerical 
comparison with data.

\begin{figure*}[t!]\centering
\includegraphics[width=.43\textwidth,clip=]{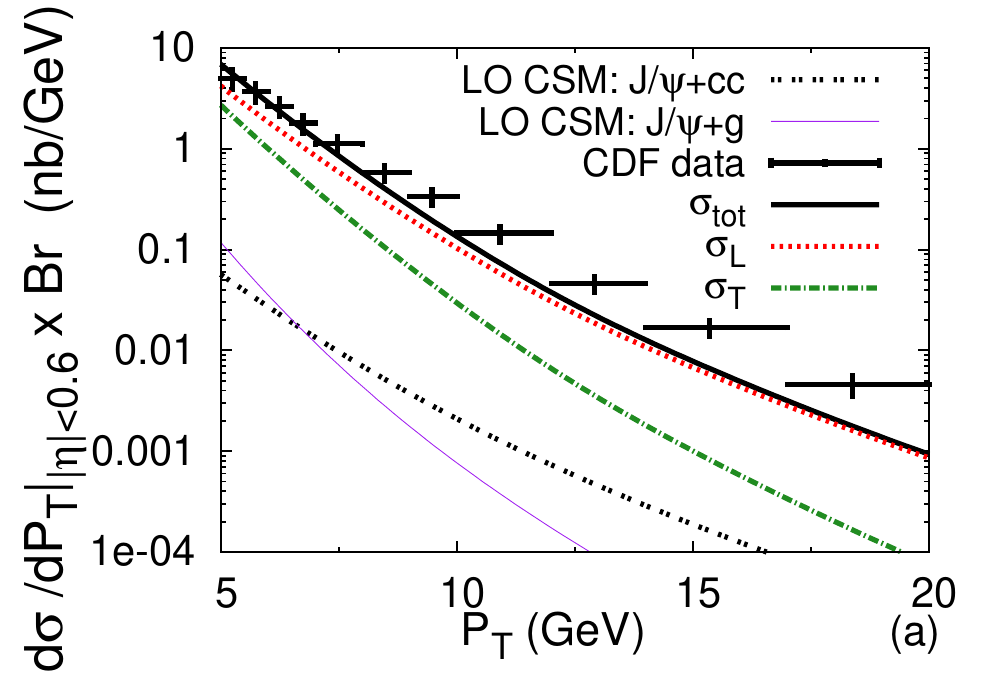}
\qquad
\includegraphics[width=.43\textwidth,clip=]{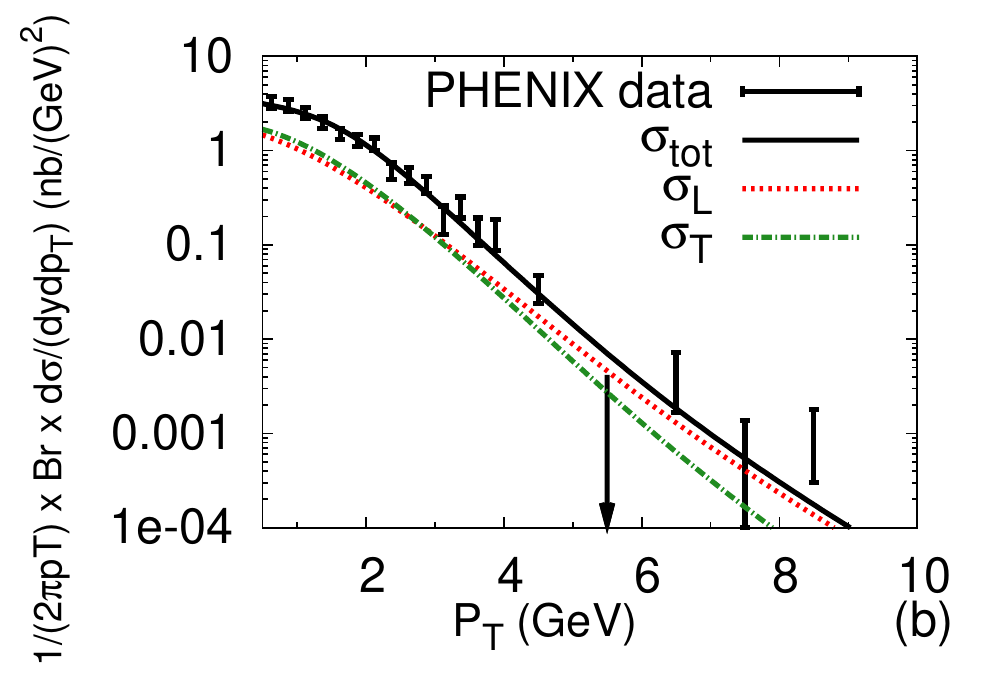}
\caption{(a) Comparison between polarised ($\sigma_T$ and $\sigma_L$) and
unpolarised ($\sigma_{\text{tot}}$) cross sections [with parameters $a=4$,
$\kappa=4.5$ GeV in Eq.~(\ref{eq:interpolate})], LO CSM contributions, and CDF
experimental data~\cite{Abe:1997yz} at the Tevatron ($\sqrt{s}=1.8$ TeV,
pseudorapidity $|\eta|<0.6$); (b) Comparison between $\sigma_T$, $\sigma_L$,
$\sigma_{\text{tot}}$ and PHENIX data~\cite{Adare:2006kf} at RHIC
($\sqrt{s}=200$ GeV, rapidity $|y|<0.35$). Taken 
from~\cite{Haberzettl:2007kj}.} \label{fig:cross-sections}
\end{figure*}

In order to obtain a correct scaling at large $P_T$ and a behaviour close to
 the minimal substitution at low $P_T$, we have 
chosen~\cite{Haberzettl:2007kj,Lansberg:2008jn}
\begin{equation}
F(c_1,c_2,q)= \Gamma_0-h(c_1\cdot
c_2)\frac{\left(\Gamma_0-\Gamma_1\right)\left(\Gamma_0-\Gamma_2\right)}{\Gamma_0}~,
\end{equation}
where the (crossing-symmetric) function $h(c_1\cdot c_2)$  rises to become
unity for large relative momentum. The phenomenological choice for the
interpolating function $h$ used in our calculations is
\begin{equation}
h(c_1\cdot c_2)= 1- a\frac{\kappa^2}{\kappa^2-(c_1\cdot c_2+m^2)}~,
 \label{eq:interpolate}
\end{equation}
with two parameters, $a$ and $\kappa$.

Figure~\ref{fig:cross-sections}(a) shows our results for $\sqrt{s}=1.8$ TeV 
in the pseudorapidity range $|\eta|<0.6$  with parameter values
$a=4$ and $\kappa=4.5$\,GeV fixed to reproduce, up to $P_T\simeq$ 10 GeV, 
the cross-section measurement of direct $J/\psi$ by
CDF~\cite{Abe:1997yz}, the usual LO CSM from $gg\to ~J/\psi\,
g$~\cite{CSM_hadron} and LO CSM from $gg\to ~J/\psi\, c
\bar{c}$~\cite{Artoisenet:2007xi}. Our results fit well the CDF
data up to about $P_T=10$ GeV. At higher $P_T$, our curve falls below the data
as expected from the genuine ${1}/{P^8_T}$ scaling of a LO box diagram 
(see the discussions of the previous section).

It is interesting to note the different $P_T$ behaviours of $\sigma_T$ and
$\sigma_L$ leading to a dominance of the latter at large $P_T$ and a negative
value for the polarisation $\alpha$~\cite{Haberzettl:2007kj} at mid and large
 $P_T$. Figure~\ref{fig:cross-sections}(b) shows
our results at $\sqrt{s}=200$\,GeV, still with  $a=4$ and $\kappa=4.5$\,GeV,
compared with the PHENIX data~\cite{Adare:2006kf}.

Through this first evaluation of the $s$-channel cut contribution
to the imaginary part of the production amplitude, 
incorporating low- and large-energy constraints as well as 
gauge invariance, we have shown that this cut can be significant. It is even
possible to obtain a very good fit of the data from CDF at mid $P_T$ 
by proper choices of the 
parametres of our 4-point function. With the same parametres, we obtained
an excellent description of the data taken at RHIC and down to very
low $P_T$ without re-summing initial-state gluon contributions. The $s$-channel
cut indeed has a threshold at low $\hat s$ (thus low $P_T$) which corresponds 
to the energy needed to put the two $c$-quark on-shell.

Now that we have seen that the $s$-channel cut matters at low- and mid-$P_T$, 
it is necessary to have in the future a first evaluation of the contribution
 of the real part itself. On the other hand, we can start testing our 
parametrisation of the 4-point function, in photoproduction 
for instance, or in any other process involving a final state gluon.

\section{Other theoretical advances }
\label{sec:th-adv-in-pp}

Beside the theoretical advances concerning QCD corrections
and the inclusion of the $s$-channel cut contribution discussed
in the previous sections, several  interesting theoretical results have been 
obtained in the recent years. Let us review some of the most significant ones
 briefly.

On the side of NRQCD, Nayak, Qiu and Sterman provided an up-to-date 
proof~\cite{nayak1} of NRQCD factorisation holding true at any order
in $v$ in the gluon-fragmentation channel. They showed that new
definitions of NRQCD matrix elements incorporating QCD Wilson lines
were to be used, but that this was not to affect the existing phenomenological 
studies.

Last year, Collins and Qiu~\cite{Collins:2007nk} showed that in
general the $k_T$-factorisation theorem does not hold in production of
high-transverse-momentum particles in hadron-collision processes, and
therefore also for $\psi$ and $\Upsilon$. This is unfortunate since
many studies~\cite{Hagler:2000dda,Hagler:2000dd,Hagler:2000eu,Yuan:2000cp,Yuan:2000qe,Baranov:2002cf,Saleev:2003ys,Kniehl:2006vm,Baranov:2007ay,Baranov:2007dw}, predicting mostly longitudinal yields and
smaller CO LDMEs, in better agreement with the idea of LDME universality, were
based on the hypothesis of such a factorisation in hadroproduction.

Besides, the $c$- and $b$-fragmentation approximation was shown to fail 
for the  $P_T$ ranges accessible in experiments for quarkonium hadroproduction.
By studying the entire set of diagrams contributing to $\psi$ and $\Upsilon$ 
production in association with a heavy-quark pair of the same flavour, we have
 shown~\cite{Artoisenet:2007xi} that the full contribution was significantly 
above (typically of a factor of 3) that obtained in the fragmentation 
approximation. A precision of 10\% accuracy, say, can only be obtained at
very large $P_T$ : $P_T \gtrsim 60$~GeV for $\psi$ and $P_T \gtrsim 100$~GeV 
for $\Upsilon$.  Note that the same observation was previously made for the 
process $\gamma \gamma \to J/\psi c \bar{c}$~\cite{Qiao:2003ba} and also for 
the  $B_c^*$ hadroproduction, for which it was noticed that the fragmentation 
approximation was not reliable at the 
Tevatron~\cite{Chang:1994aw,Berezhnoi:1996ks}.

Moreover, still in double-heavy-quark-pair production, the notion of 
colour-transfer enhancement was introduced by Nayak, Qiu and 
Sterman~\cite{nayak2}. If three out of the four heavy quarks 
are produced  with similar velocities, then there is the possibility that 
colour exchanges within this 3-quark system could turn  CO configurations 
into CS ones, thus could effectively increase the rate of production of CS 
pairs. They finally discussed the introduction of specific new 3-quark 
operators --beyond the usual ones of NRQCD-- necessary to deal with such 
an issue.
A study of the colour-transfer effects in hadroproduction is still awaited for.

\section{Associated production channels}

As previously discussed, the results of QCD corrections for $\Upsilon$ 
production seem to indicate that the CS transitions are dominant. Eventually, 
this should put an end to the controversy related to $\Upsilon$ production. 
Contrariwise, the situation remains unclear for the charmonium case. NLO 
corrections complemented by some dominant $\alpha_S^5$ contributions
are large and seem to bring the prediction for the CS transitions very close 
to the data in the $\psi'$ case for instance (see \cf{fig:dsdpt_NNLO} (b)). 
Yet, theoretical uncertainties remain large and there seems to be some space 
left for CO transitions. Careful comparisons are still therefore due with 
polarisation observables. In this case, the theoretical uncertainties would 
certainly be competitive with experimental ones, for instance on prompt 
$J/\psi$ yield~\cite{Artoisenet:2008bis}. This requires however some knowledge
 on the QCD corrections to the $P$-wave CS yield. For the time being, nothing
 is known on this side.

It is therefore vital in order to progress in the understanding of
the mechanisms responsible for heavy quarkonium production to introduce, 
compute and measure new observables. One of those is the hadronic activity 
around the quarkonium~\cite{Kraan:2008hb}. Historically, UA1 compared their 
charged-track distributions with Monte Carlo simulations for a $J/\psi$ coming 
from a $B$ and  a $J/\psi$ coming from a 
$\chi_c$~\cite{Albajar:1987ke,Albajar:1990hf}. At that time  $\chi_c$ 
feed-down was still expected to be the major source of prompt $J/\psi$. 
Following either the idea of CO transitions or of CS transitions at 
higher-orders, we however  expect now more complex distributions
even for the prompt yield. It is therefore not clear if such methods
are suitable to size up the $B$-feeddown otherwise than with the measurements
of a displaced vertex typical of  a $B$ decay.

We therefore urgently need observables rather easy to predict and likely to 
test the many production models 
available~\cite{Lansberg:2006dh,Brambilla:2004wf}. We argue here that the 
study of associated production channels, first in \pp\ collisions, then in 
\pA\ and \AaAa, fills both these requirements. By associated production 
channels, we refer to $\psi + c \bar c$ and $\Upsilon + b \bar b$.

A further motivation for such studies is that similar studies carried at 
$B$-factories showed an amazingly large fraction of $J/\psi$ produced
in association with  another $c \bar c$ pair. Indeed, 
the Belle collaboration first found \cite{Abe:2002rb} $\frac{\sigma\,(e^+ e^- 
\to J/\psi +c \bar c)}{\sigma\,(e^+ e^- \to J/\psi +X)}$
to be $0.59^{-0.13}_{+0.15}\pm 0.12$. Thereafter, the analysis was 
improved and they obtained~\cite{Uglov:2004xa} 

\eqs{\frac{\sigma\,(e^+ e^- \to J/\psi +c \bar c)}{\sigma\,(e^+ e^- \to 
J/\psi +X)}&= 0.82 \pm 0.15 \pm 0.14,\\
&> 0.48 \hbox{ at 95\% CL}.}
 
Whether or not such a high fraction holds for hadroproduction as well, is 
a question which remains unanswered. Analyses at the Tevatron 
(CDF and $D\emptyset$) and at RHIC (PHENIX and STAR) are already possible. 
As computed in~\cite{Artoisenet:2007xi} 
for the RUN2 at the Tevatron at
$\sqrt{s}=1.96$ TeV, the integrated cross-section are significant :

\eqs{\sigma(J/\psi +c \bar c) \times {\cal B} (\ell^+\ell^-)& \simeq 1~ 
\hbox{nb}\\
\sigma(\Upsilon +b \bar b)\times {\cal B} (\ell^+\ell^-)& \simeq  1~ 
\hbox{pb}}

As an illustration of the potentialities at RHIC, we chose to display 
in~\cf{fig:associated-sig-STAR}  the differential cross section for 
$pp \to J/\psi +c \bar c$  computed for the STAR kinematics. Such studies 
could for instance be carried out by STAR in the next run with an integrated 
luminosities of around 50 pb$^{-1}$ if dedicated  triggers are 
available~\cite{calderon}. 

\begin{figure}[h!]\centering
\includegraphics[width=.99\columnwidth]{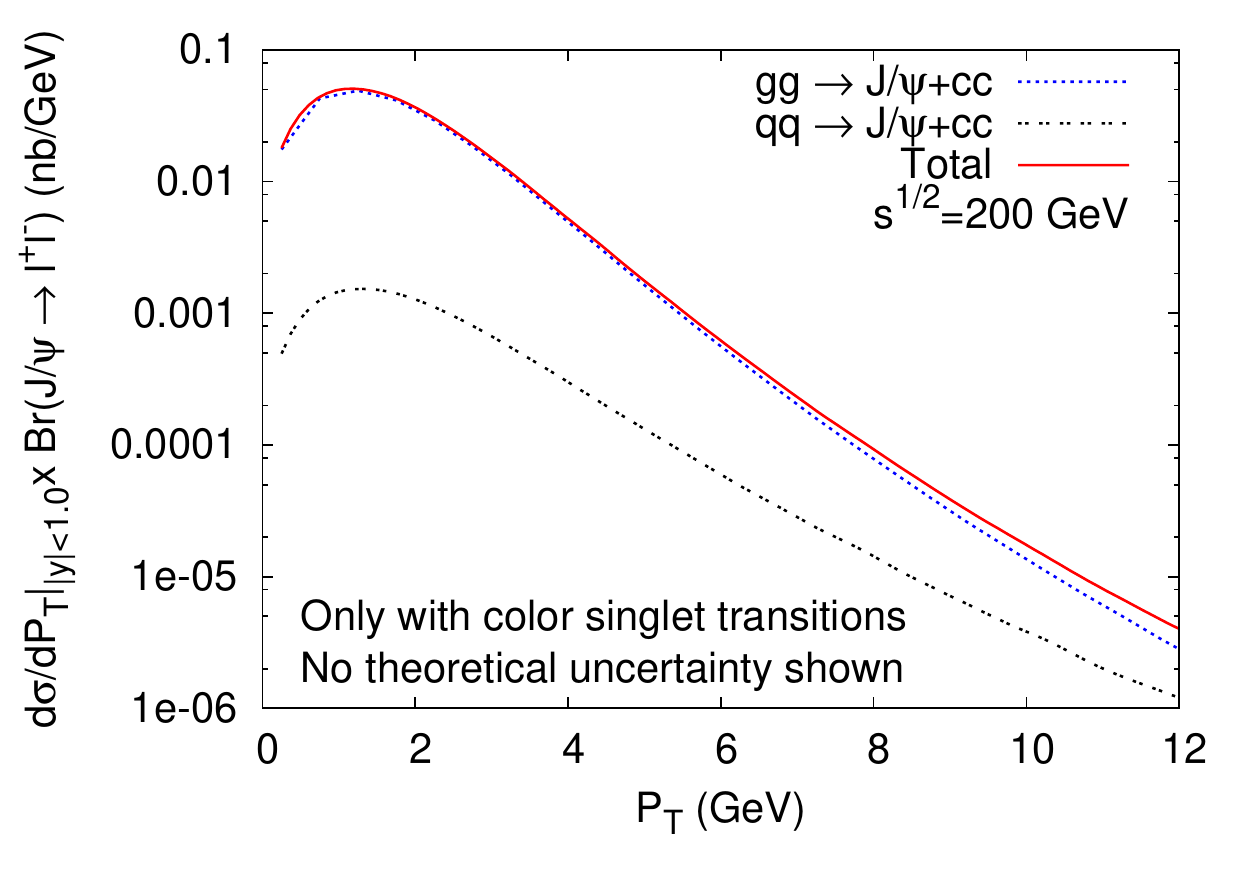}
\caption{Differential cross section for $pp \to J/\psi +c \bar c$ as function
 of the $J/\psi$ transverse momentum $P_T$ for the STAR kinematics 
($\sqrt{s}=200$ GeV, $|y|\leq 1.0$).} \label{fig:associated-sig-STAR}
\end{figure} 

Without taking into account the likely reduction of the CO LDMEs 
induced by  the QCD corrections mentioned in the previous sections, 
the integrated cross sections were found in~\cite{Artoisenet:2008tc} to be 
dominated by the CS part, similarly to the differential cross 
section in $P_T$ up to at least 5~GeV for $\psi$ and 10~GeV for $\Upsilon$. 
In other words, such observables can be thought of as a test of the CS
 contribution, for the first time since the introduction of the idea that 
CO transitions would be the dominant mechanism responsible for quarkonium 
production at high transverse momentum. If the effect of CO transitions is 
confirmed to be negligible for the $\Upsilon$,  the $\Upsilon$ produced in 
association with a $b \bar b$ pair are predicted to be strictly unpolarised, 
for any $P_T$ (see \cf{fig:associated-pol}) for the LHC.

\begin{figure}
\includegraphics[width=\columnwidth]{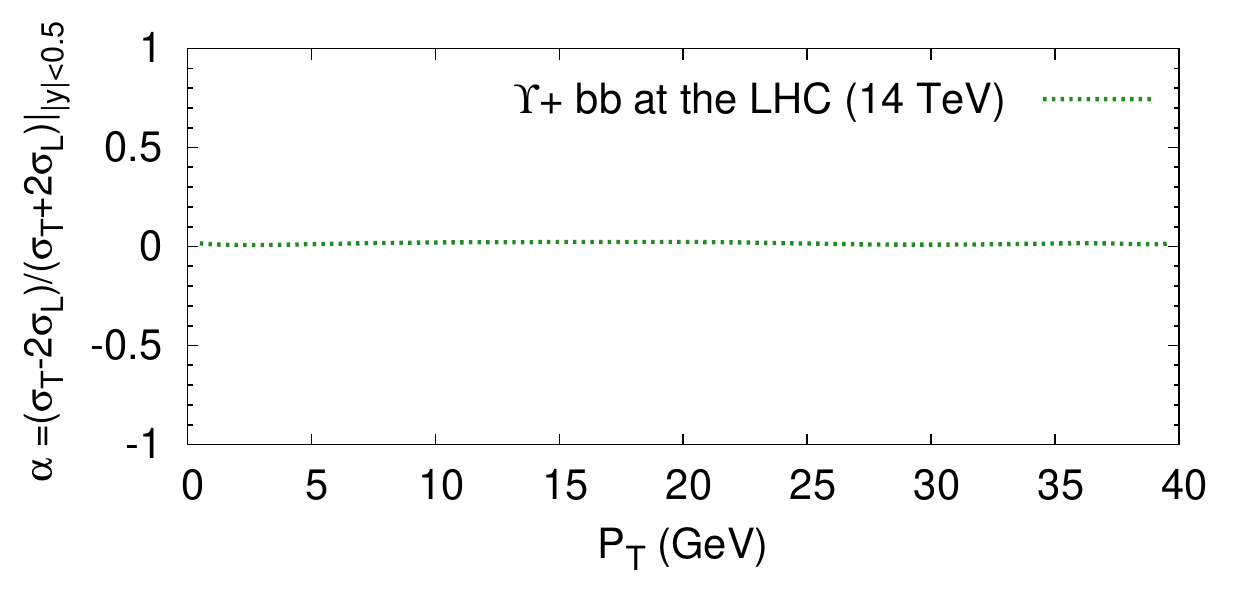}
\caption{Polarisation of an $\Upsilon$ produced in association with a 
$b\bar b$ pair at the LHC
for $\sqrt{s}=14$ TeV for $|y|\leq 0.5$. } \label{fig:associated-pol}
\end{figure} 

Beside the property of discriminating between the CO and the CS 
transitions, the yield of $\psi$ in association with $c\bar c$ should show 
an {\it a priori} completely different sensitivity to the $\chi_c$ feed-down 
than the inclusive yield. The same holds for $\Upsilon$ with $b\bar b$ with 
the $\chi_b$ feed-down. To what concerns CS transitions, the $P$-wave yield is
expected to be smaller than the $S$-wave one, since they are being suppressed
by powers of the relative velocity $v$ and here there is no extra gluon 
needed to be attached to the heavy-quark loop to produce $\psi$ 
(or $\Upsilon$) compared to $P$-waves as it is the case in the inclusive case.

For the CO transitions, associated production $\chi_c+c\bar c$ can occur via 
the process $gg\to gg$ for which the two final-state gluons split into a 
$c \bar c$ pair, one of them hadronising into a  $\chi_c$ via the CO mechanism. 
This contribution is certainly suppressed up to $P_T \simeq 20$ GeV. For larger
$P_T$, a dedicated calculation is needed. However, this mechanism would be very
easily disentangled from the CS contributions since both $c$ quarks are 
necessarily emitted back to back to the  $\chi_c$ and thus to the $J/\psi$.

Concerning the non-prompt signal, it would originate as usual from  
$gg\to b\bar b$, where one $b$ quark hadronises in $\psi$. Usually, this 
hadronisation of the $b$ produces the $\psi$ with light quarks only. This means
than we have  one single $c$ quark in the event. It is produced from the decay
of the recoiling $b$ quark and is therefore back to back to the $\psi$. The 
non-prompt signal would then be simply cut down by searching for a $D$ meson
near the $\psi$. Now it can happen that the hadronisation of the $b$ produces 
the $\psi$ and a $D$ meson. In this case, kinematical cuts would not help to 
suppress the non-prompt yield. Fortunately, this is {\it a priori} suppressed 
compared to the first case and even more than the direct yield since there is 
here no gain in the $P_T$ dependence since both  $gg\to b\bar b$ and 
$gg\to \psi +c \bar c$ cross sections scale like $P_T^{-4}$. A cross check by 
sizing up the non-prompt yield with a displaced vertex measurement would be 
anyhow surely instructive.

Let us also mention that associated production has also been studied in 
direct $\gamma\gamma$ collisions in Ultra-peripheral collision 
(UPC)~\cite{Klasen:2008mh}. At least for direct
$\gamma\gamma$ collisions, associated production is the dominant 
contribution to the inclusive rate for $P_T \geq 2$~GeV/$c$.

To conclude,  studies can be carried on by
detecting either the ``near'' or ``away'' heavy-quark with respect to
the quarkonia. There are of course different way to detect the $D$,
$B$, or a $b$-jet, ranging from the use of a displaced vertex 
to the detection of their decay in $e$ or $\mu$. As discussed above, this has 
to be considered by also taking into account the different backgrounds. 
The forthcoming Quarkonium-event-generator Madonia~2~\cite{madonia2} will surely
be of a great help to achieve this task. 
In any case, we hope that such measurements would provide with clear
information on the mechanisms at work in quarkonium production\footnote{Note also that
NLO QCD corrections have recently been computed for the production of a  $J/\psi$
and $\Upsilon$ in association with a photon~\cite{Li:2008ym}. An experimental 
study of such process could be interesting as well.}.

\section{Conclusion}

Recently significant progresses have been made in the evaluation of the
QCD corrections to quarkonium production. The situation sounds now rather clear 
for the {\it bottomonia} where an agreement has been eventually obtained using 
only CS channels when dominant $\alpha_S^5$ contributions are incorporated. The
polarisation predictions for the latter cases seem also quite encouraging 
considering CDF~\cite{Acosta:2001gv} and $D\emptyset$~\cite{D0:2008za}
measurements. Yet, confirmations are awaited for from the LHC.  

On the other hand, those $\alpha_S^5$ contributions could be still unable to 
bring agreement with the measured $P_T$-differential cross-section of the
direct {\it charmonia}.  Dedicated further studies are however needed especially
to what concerns the feed-down from $P$-waves which is not known at NLO accuracy.
In the charmonium case, we have also seen that $s$-channel cut can bring a 
significant contribution to the cross section at low $P_T$ and hence a first 
evaluation of the real part of the production amplitude is needed.

 Additional tests are now 
undoubtedly needed beyond the {\it mere} measurements of inclusive
cross section and polarisation at the LHC. For instance, the
hadroproduction of $J/\psi$ or $\Upsilon$ with a heavy-quark 
pair~\cite{Artoisenet:2007xi,Artoisenet:2008tc} appears to be a
new valuable tool to separately probe the CS contribution, at least
dominant at low-$P_T$ (below 15~GeV), as well as the study of the
hadronic activity around the quarkonium.

\section*{Acknowledgements}
The work on QCD corrections is done in collaboration with P. Artoisenet, 
J. Campbell, F. Maltoni and 
F. Tramontano~\cite{Artoisenet:2008fc,Artoisenet:2008bis} and on
the $s$-channel cut contribution with J.R. Cudell, H. Haberzettl and 
Yu.L. Kalinovsky.  We would like to  warmly thank the organisers of HP2008 
for their kind invitation to present this overview on quarkonium production.
This work is supported in part by a Francqui fellowship of the Belgian American Educational Foundation and
by the U.S. Department of Energy under contract number DE-AC02-76SF00515.

\end{document}